\documentclass[preprint2]{aastex6}
\usepackage{cancel}	
\usepackage{color}
\usepackage{affils}
\usepackage{graphicx,subfigure}
\usepackage{upgreek}

\newcommand{\kms}{km\,s$^{-1}$}

\def\M{M$_{\odot}$}

\def\Ni{$^{56}$Ni}
 \def\Co{$^{56}$Co}

 \def\Mej{$M_{\rm ej}$}

\begin{document}


\title{Superluminous supernova SN\,2015bn in the nebular phase: evidence for the engine-powered explosion of a stripped massive star}

\shortauthors{Nicholl et al.}

\shorttitle{SLSN 2015bn in the nebular phase}


\DeclareAffil{cfa}{Harvard-Smithsonian Center for Astrophysics, 60 Garden Street, Cambridge,
  Massachusetts 02138, USA; \href{mailto:matt.nicholl@cfa.harvard.edu}{matt.nicholl@cfa.harvard.edu}}
\DeclareAffil{queens}{Astrophysics Research Centre, School of Mathematics and Physics, Queens University Belfast, Belfast BT7 1NN, UK}
\DeclareAffil{nyu}{Center for Cosmology and Particle Physics, New York University, 4 Washington Place, New York, NY 10003, USA}
\DeclareAffil{mpe}{Max-Planck-Institut f{\"u}r Extraterrestrische Physik, Giessenbachstra\ss e 1, 85748, Garching, Germany}
\DeclareAffil{mpa}{Max-Planck-Institut f{\"u}r Astrophysik, Karl-Schwarzschild-Str. 1, D-85748 Garching, Germany}
\DeclareAffil{lcogt}{Las Cumbres Observatory Global Telescope, 6740 Cortona Dr, Suite 102, Goleta, CA 93111, USA}
\DeclareAffil{kitp}{Kavli Institute for Theoretical Physics, University of California, Santa Barbara, CA 93106, USA}
\DeclareAffil{soton}{School of Physics and Astronomy, University of Southampton, Southampton, SO17 1BJ, UK}
\DeclareAffil{hawaii}{Institute for Astronomy, University of Hawaii at Manoa, Honolulu, HI 96822, USA}
\DeclareAffil{ohio}{Astrophysical Institute, Department of Physics and Astronomy, 251B Clippinger Lab, Ohio University, Athens, OH 45701, USA}
\DeclareAffil{cambridge}{Institute of Astronomy, University of Cambridge, Madingley Road, Cambridge, CB3 0HA}
\DeclareAffil{weizmann}{Benoziyo Center for Astrophysics, Weizmann Institute of Science, Rehovot 76100, Israel}
\DeclareAffil{millennium}{Millennium Institute of Astrophysics, Vicu\~{n}a Mackenna 4860, 7820436 Macul, Santiago, Chile}
\DeclareAffil{chile}{Departamento de Astronom\'ia, Universidad de Chile, Camino El Observatorio 1515, Las Condes, Santiago, Chile}
\DeclareAffil{tuorla}{Tuorla Observatory, Department of Physics and Astronomy, University of Turku, V\"ais\"al\"antie 20, FI-21500 Piikkiö, Finland}
\DeclareAffil{osu}{Department of Astronomy, The Ohio State University, 140 West 18th Avenue, Columbus, OH 43210, USA}
\DeclareAffil{ccapp}{Center for Cosmology and AstroParticle Physics (CCAPP), The Ohio State University, 191 W. Woodruff Ave., Columbus, OH 43210, USA}
\DeclareAffil{ucsb}{Department of Physics, University of California, Santa Barbara, Broida Hall, Mail Code 9530, Santa Barbara, CA 93106-9530, USA}
\DeclareAffil{taiwan}{Institute of Astronomy, National Central University, Chung-Li 32054, Taiwan}
\DeclareAffil{columbia}{Columbia Astrophysics Laboratory, Columbia University, New York, NY 10027, USA}
\DeclareAffil{sorbonne}{Sorbonne Universit\'es, UPMC, Paris VI,UMR 7585, LPNHE, F-75005, Paris, France}
\DeclareAffil{cnrs}{CNRS, UMR 7585, Laboratoire de Physique Nucleaire et des Hautes Energies, 4 place Jussieu, 75005 Paris, France}
\DeclareAffil{not}{Nordic Optical Telescope, Apartado 474, E-38700 Santa Cruz de La Palma, Spain}
\DeclareAffil{carnegie}{Carnegie Observatories, 813 Santa Barbara Street, Pasadena, CA 91101, USA}
\DeclareAffil{hcpf}{Hubble, Carnegie-Princeton Fellow}
\DeclareAffil{davis}{Department of Physics, University of California, Davis, CA 95616, USA}
\DeclareAffil{mit}{Kavli Institute for Astrophysics and Space Research, Massachusetts Institute of Technology, 77 Massachusetts Avenue, Cambridge, MA 02139}
\DeclareAffil{caltech}{Astronomy Department, California Institute of Technology, Pasadena, California 91125, USA}
\DeclareAffil{jmu}{Astrophysics Research Institute, Liverpool John Moores University, IC2, Liverpool Science Park, 146 Brownlow Hill, Liverpool L3 5RF, UK}
\DeclareAffil{eso}{European Southern Observatory, Karl-Schwarzschild-Str. 2, D-85748 Garching, Germany}

\affilauthorlist{M.~Nicholl\affils{cfa},
E.~Berger\affils{cfa},
R.~Margutti\affils{nyu},
R.~Chornock\affils{ohio},
P.~K.~Blanchard\affils{cfa},
A.~Jerkstrand\affils{queens},
S.~J.~Smartt\affils{queens},
I.~Arcavi\affils{lcogt},
P.~Challis\affils{cfa},
K.~C.~Chambers\affils{hawaii},
T.-W.~Chen\affils{mpe},
P.~S.~Cowperthwaite\affils{cfa},
A.~Gal-Yam\affils{weizmann},
G.~Hosseinzadeh\affils{lcogt},
D.~A.~Howell\affils{lcogt},
C.~Inserra\affils{queens},
E.~Kankare\affils{queens},
E.~A.~Magnier\affils{hawaii},
K.~Maguire\affils{queens},
P.~A.~Mazzali\affils{jmu,mpa},
C.~McCully\affils{lcogt},
D.~Milisavljevic\affils{cfa},
K.~W.~Smith\affils{queens},
S.~Taubenberger\affils{mpa,eso},
S.~Valenti\affils{davis},
R.~J.~Wainscoat\affils{hawaii},
O.~Yaron\affils{weizmann},
D.~R.~Young\affils{queens}
}

\begin{abstract}

We present nebular-phase imaging and spectroscopy for the hydrogen-poor superluminous supernova SN\,2015bn, at redshift $z=0.1136$, spanning +250--400\,d after maximum light. The light curve exhibits a steepening in the decline rate from 1.4\,mag\,(100\,d)$^{-1}$ to 1.7\,mag\,(100\,d)$^{-1}$, suggestive of a significant decrease in the opacity. This change is accompanied by a transition from a blue continuum superposed with photospheric absorption lines to a nebular spectrum dominated by emission lines of oxygen, calcium and magnesium. There are no obvious signatures of circumstellar interaction or large \Ni~mass. We show that the spectrum at +400\,d is virtually identical to a number of energetic Type Ic supernovae such as SN\,1997dq, SN\,2012au, and SN\,1998bw, indicating similar core conditions and strengthening the link between `hypernovae'/long gamma-ray bursts and superluminous supernovae. A single explosion mechanism may unify these events that span absolute magnitudes of $-22 < M_B < -17$. Both the light curve and spectrum of SN\,2015bn are consistent with an engine-driven explosion ejecting $7-30$\,\M~of oxygen-dominated ejecta (for reasonable choices in temperature and opacity). A strong and relatively narrow \ion{O}{1}\,$\lambda$7774 line, seen in a number of these energetic events but not in normal supernovae, may point to an inner shell that is the signature of a central engine.

\end{abstract}


\section{Introduction}\label{sec:intro}

Superluminous supernovae (SLSNe) are the brightest explosions in the Universe at optical wavelengths \citep[see][]{gal2012}, but long evaded detection due to their rarity and preference for low-metallicity dwarf galaxies \citep[e.g.][]{chen2013,lun2014,lel2015}. Since they were first recognised \citep{qui2011}, the number of known SLSNe---almost invariably discovered by untargeted transient surveys---has grown to several tens \citep{nic2015b}.

The spectra of Type I SLSNe show neither hydrogen nor low-velocity lines to clearly indicate interaction with circumstellar medium (CSM). A few events have exhibited hydrogen emission after several hundred days, but this is not directly related to their power source at peak \citep{yan2015}. Early spectra are instead dominated by \ion{O}{2} absorption and blue continua indicating hot photospheres. After sufficient cooling, they evolve to resemble normal and broad-lined Type Ic supernovae (SNe) of more typical luminosity \citep{pas2010}.

An appealing model to power SLSNe I is a `central engine' \citep[such as a millisecond magnetar;][]{shk1976,dunc1992,kas2010,woo2010} with a timescale similar to the diffusion time, thus thermalising the input without significant adiabatic dilution \citep{met2015}. Evidence for this scenario includes light curve fitting \citep{ins2013,nic2013}, similar host galaxies to long gamma-ray bursts \citep[LGRBs;][]{lun2014,per2016}, and large kinetic energies \citep{nic2015b}. Recently, the ultra-long GRB\,111209A was associated with a very bright (though not quite superluminous) supernova, SN\,2011kl \citep{gre2015}.

A crucial test for any model is provided by observations hundreds of days after explosion when the ejecta are largely transparent, offering direct constraints on the internal conditions and distribution of material. Only 2 SLSNe have spectra later than 300\,d after maximum light: SN\,2007bi \citep{gal2009} and PTF12dam \citep{chen2014}. These have been of low signal-to-noise or strongly contaminated by the host \citep{jer2016}. In this letter we present deep photometric and spectroscopic observations of SN\,2015bn---the best-observed SLSN to date \citep{nic2016b}---at 250 to 400\,d after peak\footnote{all phases in rest-frame}, and show that the nebular spectrum connects SLSNe to other energetic and potentially engine-powered explosions spanning a wide range in luminosity.

\section{Observations}\label{sec:obs}

\begin{figure*}
\centering
\includegraphics[width=9.5cm]{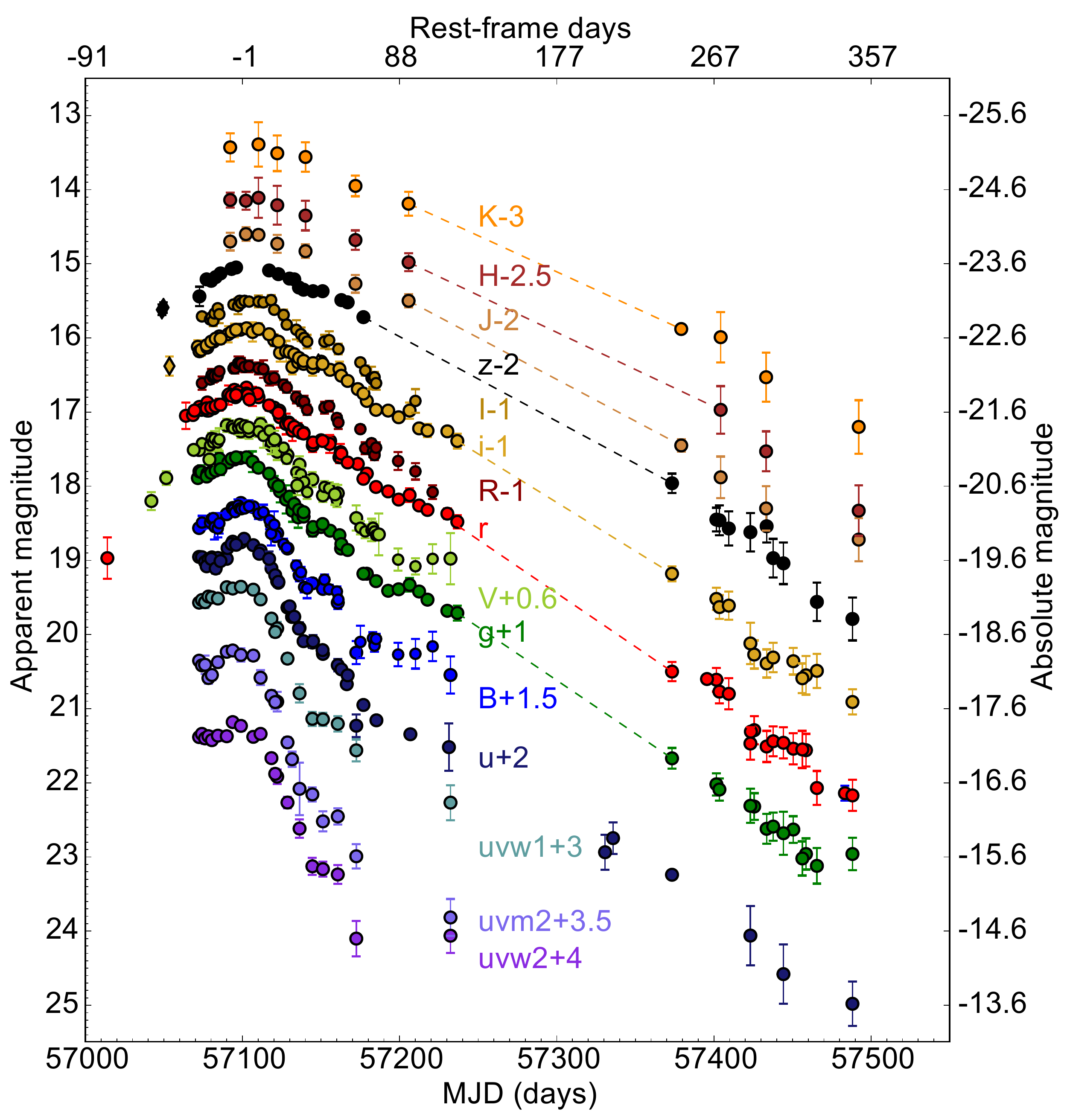}
\includegraphics[width=7.8cm]{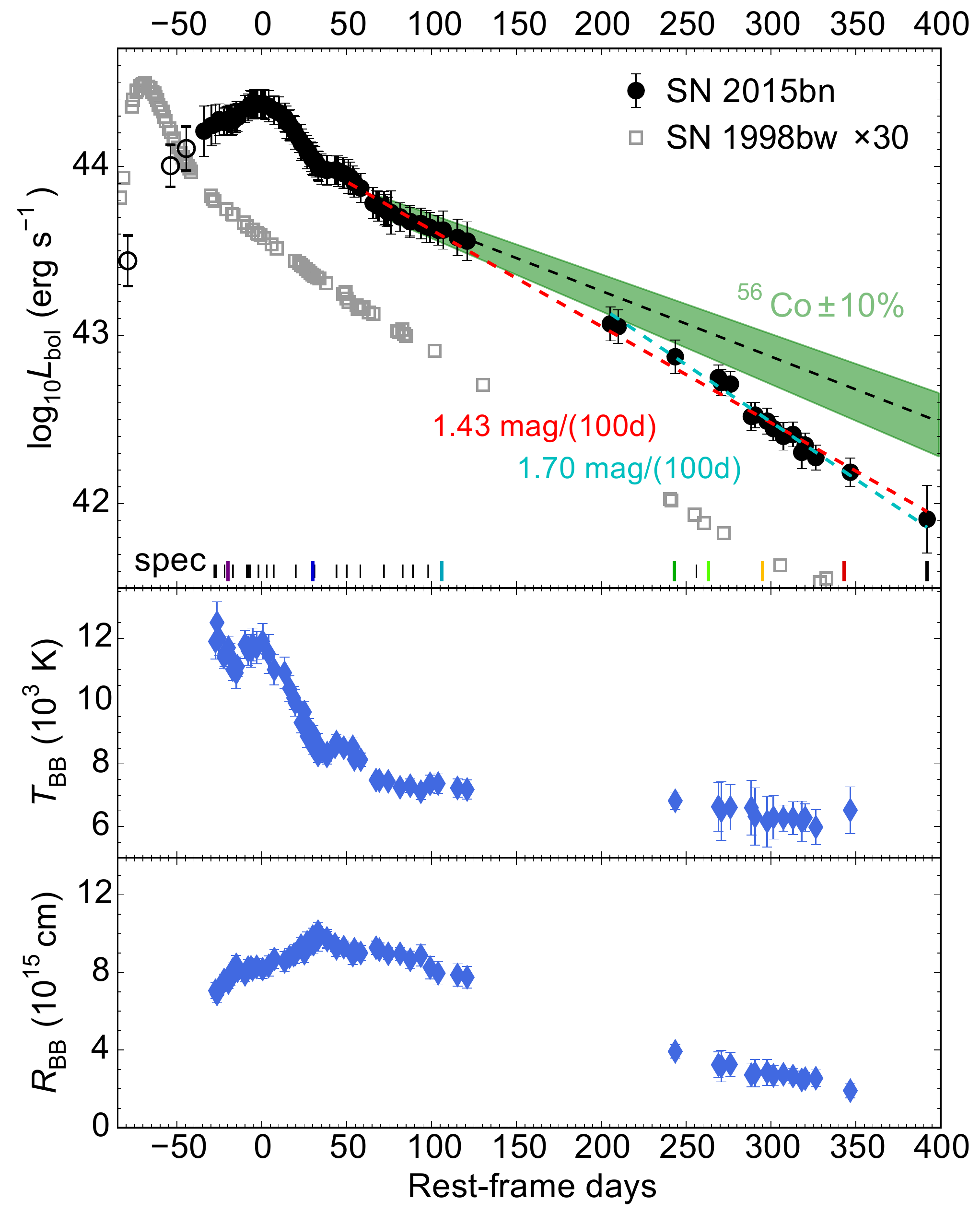}
\figcaption{Left: Multicolour light curves up to +400\,d, showing a slow decline. No other SLSN has such wide wavelength coverage at this phase. Right: Bolometric light curve with blackbody colour temperature and radius. Open symbols estimated from single-filter photometry. Final point is from integrating our GMOS spectrum. Epochs with spectroscopy are marked.
\label{fig:phot}}
\end{figure*}

\subsection{New data}

Observations of SN\,2015bn from earlier than +250\,d after peak were presented and analysed by \citet{nic2016b}. New imaging was obtained with EFOSC2 and SOFI on the New Technology Telescope \citep[through PESSTO;][]{sma2015}, the Las Cumbres Observatory Global Telescope Network (LCOGT), and the Pan-STARRS Survey for Transients \citep{hub2015}. Spectra were taken with IMACS on the Magellan Baade telescope and GMOS on Gemini North. The GMOS spectrum was reduced using the \textsc{gemini} package in \textsc{iraf}; other data reduction is described by \citet{nic2016b}.

At late times, the host galaxy contributes significant flux, which must be removed. For EFOSC2 images, we subtract pre-SN images from the Sloan Digital Sky Survey \citep{alam2015} using \textsc{hotpants}\footnote{http://www.astro.washington.edu/users/becker/v2.0/hotpants.html}. For LCOGT images, subtraction performed poorly due to point spread function variations. Instead, we took the average difference in flux measured before and after subtraction in the EFOSC2 images, and subtracted this from our LCOGT measurements. This worked well, as the compact host is approximately a point source. For the near-infrared, we did not have a good measurement of the host magnitudes. In this case we assumed $J=21.5$, $H=21$ and $K=20.5$ (Vega), based on the optical-NIR colours of the SLSN host galaxies studied by \citet{chen2014} and \citet{ang2016}. Note that our analysis is insensitive to precise host removal at these wavelengths, as the NIR contribution to the bolometric luminosity is $\lesssim20\%$. We make no correction for flux beyond 2.6\,$\upmu$m, as no such measurements exist for SLSNe. \citet{chen2014} discuss potential mid-infrared emission at late times, assuming a flux contribution of $\sim20\%$ for a temperature of 3000\,K. Our temperature estimates are somewhat higher (see below) and the contribution is only a few percent under the coarse assumption of a blackbody SED.

For our spectra, we subtract the host model from \citet{nic2016b} after rescaling to better match the SDSS $griz$ magnitudes (since the $u$-band detection is not robust). Data after subtraction are listed in Table \ref{tab:data}.

\begin{figure*}
\centering
\includegraphics[width=17cm]{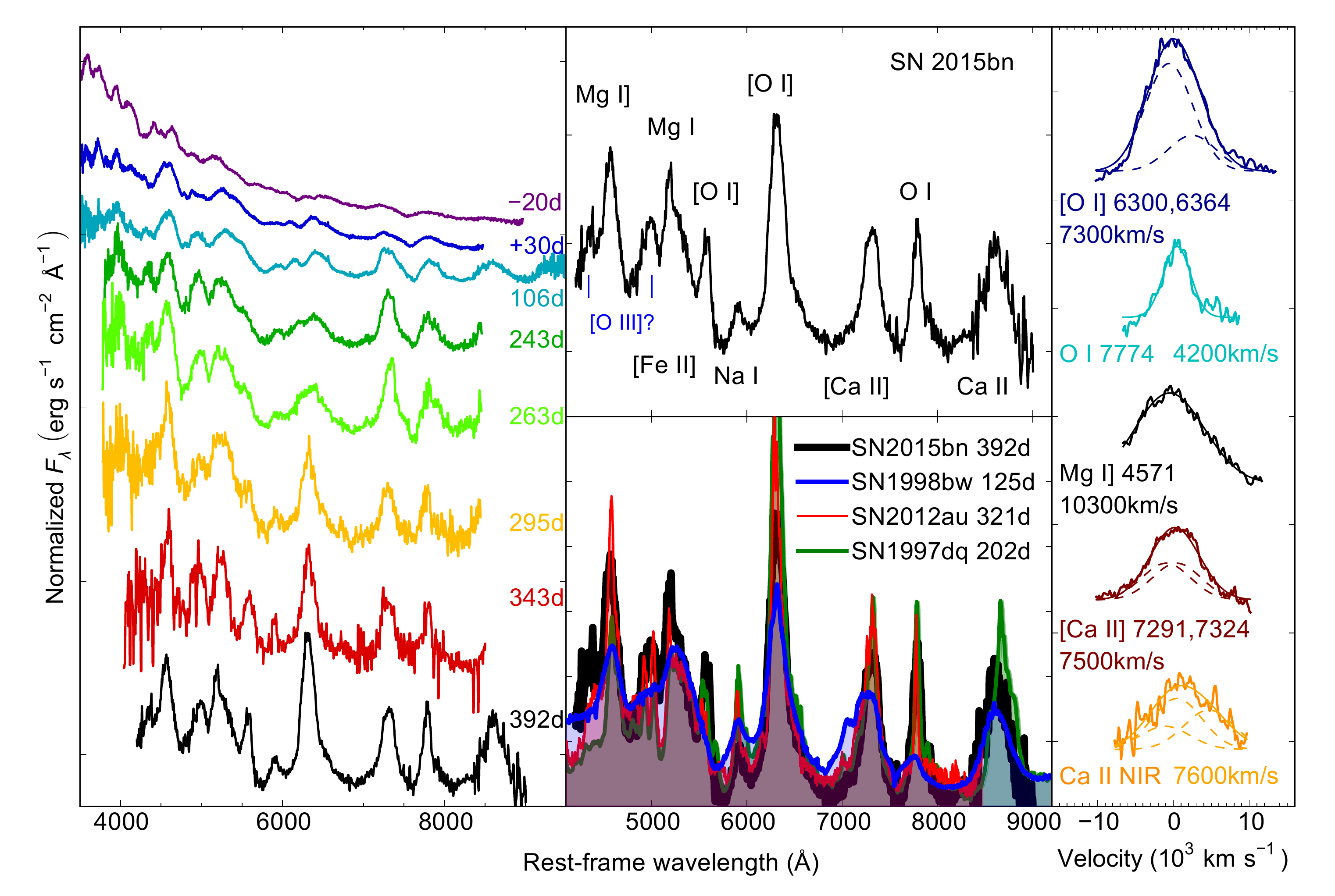}
\figcaption{Left: Spectroscopic evolution. All spectra have been normalised using the integrated flux between 4400--8000\,\AA. Middle: the GMOS spectrum at +392\,d, dominated by emission from oxygen, calcium and magnesium, is a near-perfect match to the nebular spectra of hypernovae. Right: Gaussian fits to the strongest  lines. Multiplets have been accounted for using multi-component Gaussians of the same velocity (relative strengths assume lines are optically thin). As in SN\,2012au \citep{mil2013}, \ion{O}{1}\,$\lambda7774$ exhibits a lower velocity than [\ion{O}{1}]. Note: galaxy lines have been removed for clarity.
\label{fig:spec}}
\end{figure*}

\subsection{Photometry}\label{sec:phot}

The light curves of SN\,2015bn are shown in Figure \ref{fig:phot}. Photometry spanning the full range of $u$ to $K$ (made possible by PESSTO's dense coverage with both EFOSC2 and SOFI) is unprecedented for a SLSN at such late times (\citealt{chen2014} presented $gri$ out to 400\,d for PTF12dam). The luminosity at all wavelengths shows a slow decline, with little colour evolution. We also plot the bolometric light curve, including $K$-corrections from spectra. The temperature and radius of a blackbody fit follows smoothly from those presented by \citep{nic2016b}, but in fact most of the observed radiation is not thermal and thus the late-time estimates of these properties are not reliable (section \ref{sec:spec}).

The previously-published light curve out to +250\,d appears roughly compatible with the decay rate of radioactive \Co~\citep[although we were unable to find a convincing radioactive decay model;][]{nic2016b}. However, our extended light curve shows that SN\,2015bn fades significantly faster than fully-trapped \Co~decay. Linear fits from +60 to +400\,d indicate a fading rate of 0.0143\,mag\,d$^{-1}$, compared to 0.0098 for \Co. The decline is even steeper (0.017\,mag\,d$^{-1}$) between 250--400\,d. However, the steepening could be interpreted as another undulation in the light curve, such as those around maximum light. The decline rate is actually similar to that of GRB-SN\,1998bw \citep{pat2001}, but SN\,2015bn is brighter by a factor of $\sim$150 at the same phase from explosion. The \Ni~mass in SN\,1998bw has been estimated as $\approx0.3-0.7$\,\M~\citep[e.g.][]{sol2000,maz2001}; therefore the luminosity of SN\,2015bn would require at least a few tens of solar masses of \Ni~to power the late-time light curve. However, a precise measurement is difficult as long as the light curve fades faster than \Co-decay. Applying Arnett's rule \citep{arn1982} to the light curve peaks and assuming rise times of 80\,d and 15\,d would give nickel masses of 32\,\M~and 0.4\,\M~for SN\,2015bn and SN\,1998bw, respectively. Given the long rise time of SN\,2015bn, we have included \Co~decay in this calculation.

\subsection{Spectroscopy}\label{sec:spec}

Our late-time spectroscopic series is shown in Figure \ref{fig:spec}, along with some early spectra from \citet{nic2016b}. The spectrum at +243\,d shows some emission lines---[\ion{Ca}{2}]\,$\lambda\lambda7291,7324$, \ion{Ca}{2}\,H\&K and possibly \ion{Mg}{1}]\,$\lambda4571$---but still retains a photospheric component, with a blue continuum and absorption features. In our next spectrum, 20\,d later, the \ion{Mg}{1}] line strengthens significantly and \ion{O}{1} starts to appear in emission, where we had previously detected it in absorption or P Cygni profiles. By +295\,d, we see clear oxygen emission lines: [\ion{O}{1}]\,$\lambda\lambda6300,6364$, [\ion{O}{1}]\,$\lambda5577$, and \ion{O}{1}\,$\lambda7774$, as well as \ion{Na}{1}\,D and broad features around 5000\,\AA~normally interpreted as magnesium and iron \citep[e.g.]{maz2001,jer2016b}. In particular, [\ion{O}{1}]\,$\lambda\lambda6300,6364$ increases in strength beyond +300\,d, and is by far the strongest line by our final spectrum at +392\,d. That the ejecta take $\sim$400\,d to become nebular suggests both a large mass and a persistent source of ionising radiation to maintain a photosphere to such late times. In contrast to the SLSNe studied by \citet{yan2015}, we see no hydrogen or interaction signatures. The [\ion{Ca}{2}]/[\ion{O}{1}] flux ratio is about 0.5 after $\gtrsim300$\,d, which is similar to most Type Ic SNe \citep{elm2011} but very different from massive pair-instability SN models \citep{des2013,jer2016}. We detect features consistent with [\ion{O}{3}], though not as strong as those recently seen in SLSNe PS1-14bj and LSQ14an (\citealt{lun2016,jer2016b}, C.~Inserra et al.~in prep.).

The [\ion{O}{1}]\,$\lambda\lambda6300,6364$ line is often used as a diagnostic of asphericity \citep{tau2009}. Many H-poor SNe show a double-peaked line profile suggesting a bipolar explosion viewed at a large angle from the dominant axis \citep{maz2005,mae2008}. \citet{ins2016} recently showed from spectropolarimetry that SN\,2015bn displays a dominant axis. If it is a highly asymmetric or jet-driven explosion, the single-peaked profile we observe would point to a modest inclination.

The relatively narrower width of the \ion{O}{1}\,$\lambda7774$ line suggests that the emission comes from two distinct velocity regions \citep{mil2013}; its lower velocity could indicate that it is arising deeper in the ejecta. This density-sensitive recombination line \citep[see also section \ref{sec:dis}]{mau2010} is not typically seen with such strength in normal SNe Ic, but has been detected in a number of energetic SNe Ic as well as SLSN 2007bi \citep{gal2009,mil2013}.

In fact, our nebular spectrum exhibits striking similarity to energetic SNe Ic\footnote{Data obtained via the Open Supernova Catalog \citep{gui2016}} such as SN\,1998bw \citep{pat2001}, SN\,1997dq \citep{tau2009}, and SN\,2012au \citep{mil2013}. These events are often termed `hypernovae' as their inferred kinetic energy is $\gtrsim10^{52}$\,erg: an order of magnitude larger than in normal neutrino-driven SNe, thus seeming to require an additional engine \citep{iwa1998}. The observational link between some hypernovae and LGRBs, demonstrated spectacularly by SN\,1998bw \citep{gal1998}, confirms this engine as most likely a rapidly rotating compact object: either an accreting black hole `collapsar' \citep{macf1999} or a millisecond magnetar \citep{dunc1992}.  The extraordinary similarity in nebular-phase spectra (probing the conditions of the innermost ejecta from the stellar interior) demonstrates that SLSNe and hypernovae have similar conditions in their cores,  This could indicate that their progenitors or explosion mechanisms are related, consistent with both classes occurring in similar host environments \citep{lun2014,per2016}.

\begin{figure*}
\centering
\includegraphics[width=12cm]{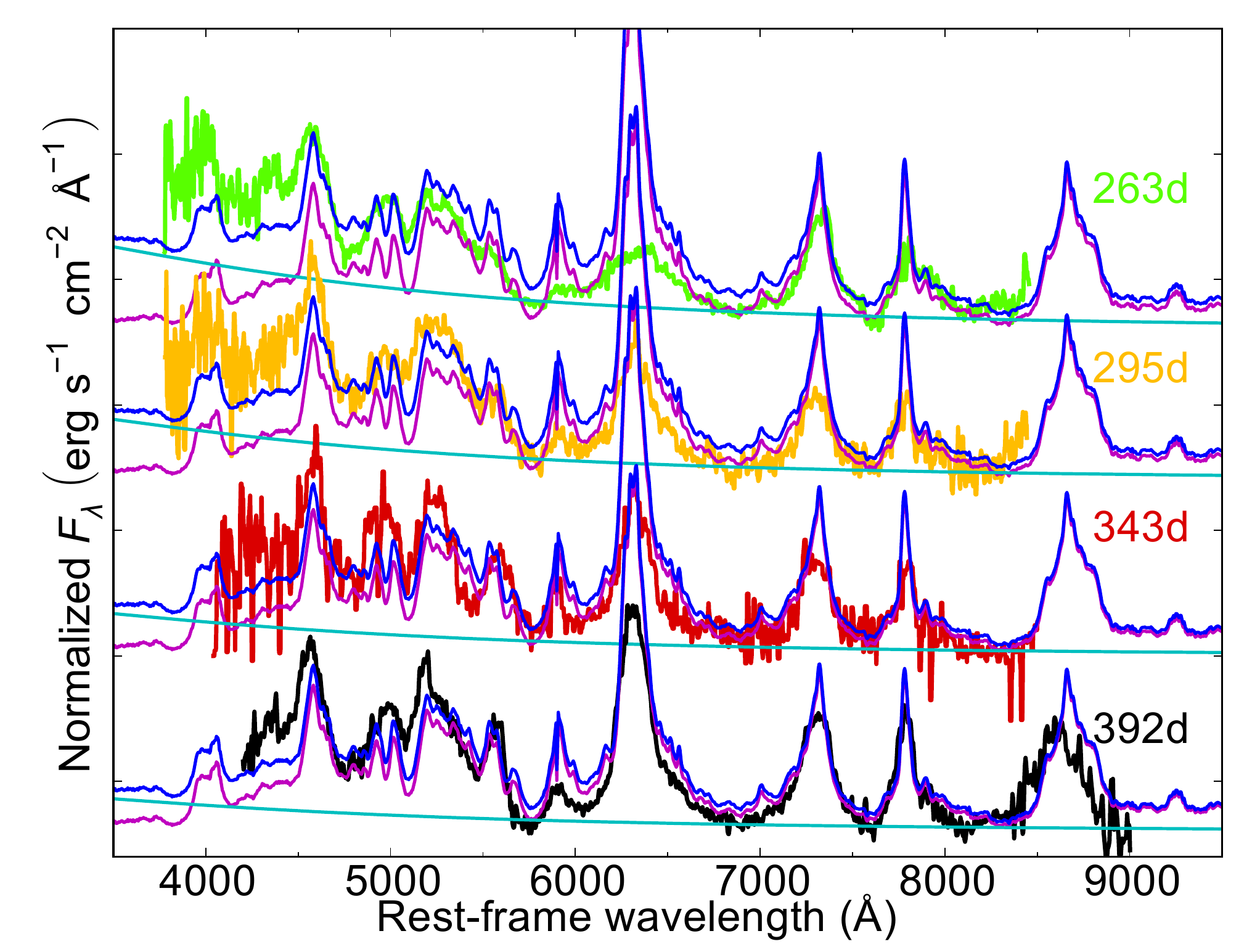}
\figcaption{Decomposition of late spectra. The overall shape is well matched by a sum of SN\,1997dq at +200\,d and a fading blue continuum that makes only a minor contribution after +300\,d.
\label{fig:cont}}
\end{figure*}

\section{Discussion}\label{sec:dis}

Given this clear link between SLSNe and hypernovae/GRB-SNe, we look to build a consistent picture of SN\,2015bn within the central-engine framework. Independent evidence for this link comes from spectropolarimetry \citep{ins2016}, which shows axisymmetry similar to GRB-SNe. While black hole accretion has also been proposed as a viable engine for SLSNe \citep{dex2013}, magnetar-powered models are likely more applicable here due to the long engine timescale required by the observations.

Although the progenitors and explosion mechanism may be similar, it seems that a different process supplies the luminosity of SN\,2015bn compared to the hypernovae \citep[which seem to be heated by \Ni, e.g.][]{cano2016b}. In section \ref{sec:phot} we saw that the larger luminosity of SN\,2015bn compared to SN\,1998bw would require $\gtrsim30$\,\M~of \Ni, but the spectroscopic similarity demonstrates that SN\,2015bn cannot have an enormously larger \Ni~fraction than the hypernovae. While 30\,\M~of \Ni~could be produced in a pair-instability SN, our spectra do not resemble pair-instability models \citep{jer2016}; nor do we see the [\ion{Fe}{3}] lines that dominate Type Ia SNe in the blue. This \Ni-mass is also comparable to our largest estimates of the total ejecta mass in SN\,2015bn (see below). With no strong signatures of CSM interaction, it seems that the engine itself most likely supplies the luminosity.

SN\,2015bn does appear to be slightly brighter in the blue than SNe\,1997dq and 2012au. This could point to a larger iron line luminosity, but we argue that it is more likely due to residual continuum flux. In Figure \ref{fig:cont}, we
show that the spectroscopic evolution from $\sim$250--400\,d can be well-reproduced by a superposition of the nebular spectrum of SN\,1997dq and a blue continuum (with the exception of [\ion{O}{1}] at earlier epochs). The models we show assume a blackbody form for the continuum; we use a temperature $\approx12000$\,K at all phases. However, this is simply intended to illustrate the blue colour, rather than assert that the continuum is necessarily thermal. By adding continuum rather than [\ion{Fe}{2}] alone, we find a good match to magnesium and calcium lines. This further removes any need for an unusually large iron mass \citep[see also][supplementary information]{nic2013}.

The continuum component fades over time as the emission lines increase in strength, and has little effect by the final spectrum. It is consistent with photospheric recession towards low mass-coordinate. 
An alternative explanation is that we are seeing leakage of radiation from the engine, although this is unlikely as most of this power should be emitted at much higher frequency \citep{met2014}. Finally, we consider the possibility that the continuum is powered by late-time circumstellar interaction. However, the spectrum is nearly identical to SN\,2012au, for which radio observations revealed a low-density CSM and only weak mass-loss in the years before explosion \citep{kam2014}. Deep radio limits on SN\,2015bn at +240\,d from \citet{nic2016b} did not find evidence for interaction (but could not definitively exclude it).

That some of the power input appears to be thermalised even at such late stages suggests a high-density inner region \citep{maz2004}. A defining prediction of magnetar-powered models is the formation of a constant-density shell due to the central overpressure from the magnetised wind \citep{kas2010,woo2010,met2014}. This shell could perhaps also help to account for the relative strength (and width) of the \ion{O}{1}\,$\lambda7774$ recombination line in SLSNe/hypernovae, if its electron density is $\gtrsim10^8$\,cm$^{-3}$ at this phase \citep{maz2004,nic2013,jer2016b}. Normal SNe Ic without an engine---or GRB-SNe, where the engine punches right through the stellar envelope rather than forming a shell---do not exhibit a strong line here.

\begin{figure*}
\centering
\includegraphics[width=8.5cm]{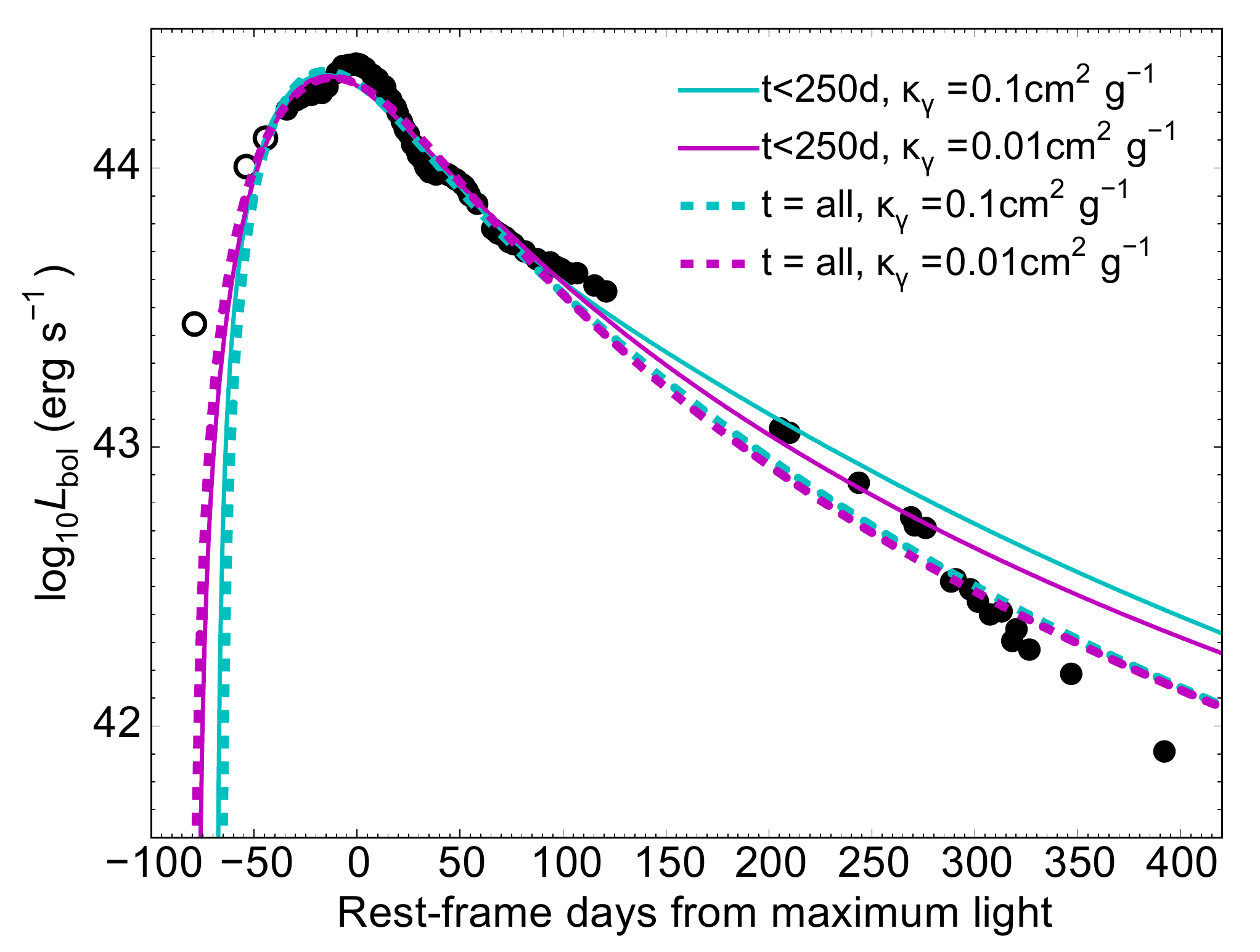}
\includegraphics[width=8.5cm]{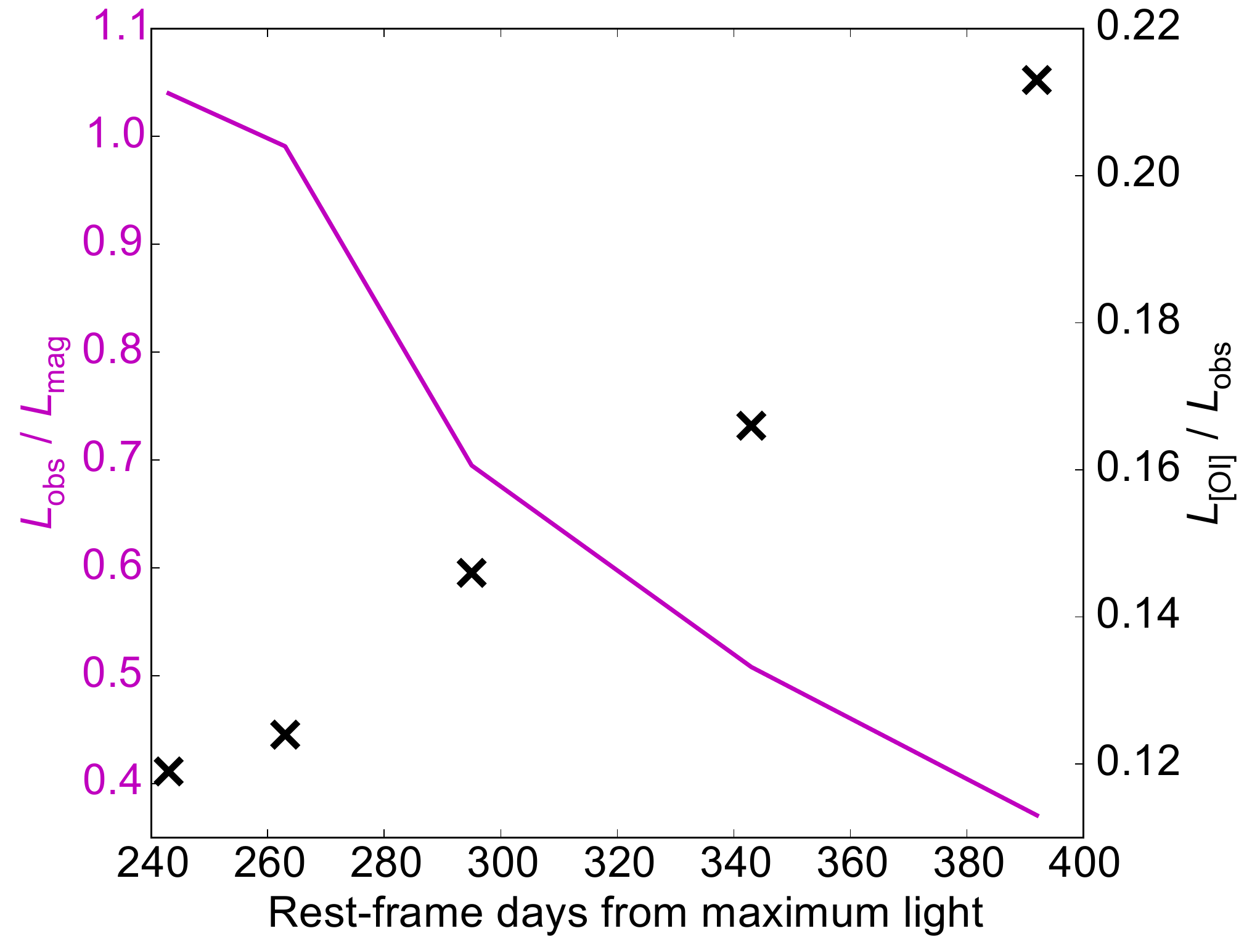}
\figcaption{Left: Magnetar-powered fits to the light curve. Parameters are (\Mej/\M, $P$/ms, $B$/$10^{14}$\,G): 8.3, 2.1, 0.9 (solid-cyan); 7.4, 1.5, 0.2 (solid-magenta); 11.9, 2.2, 0.9 (dashed-cyan); 9.0, 1.8, 0.4 (dashed-magenta). Right: Ratio of observed/model luminosity (solid-magenta model). As the ejecta become more optically thin (represented by the fraction of luminosity in the emission lines, the strongest being [\ion{O}{1}]\,$\lambda6300,6364$), the model (which assumes an optically thick photosphere) over-predicts the luminosity. The difference may manifest as X-rays/$\gamma$-rays observable directly from the magnetar.
\label{fig:mod}}
\end{figure*}

Taking equation 7 from \citet{kas2010} for the shell velocity, $v_{\rm sh}\approx (2E_{\rm sn}/M_{\rm ej})^{1/2} (1 + E_{\rm mag}/E_{\rm sn})^{(1/2)}$, with the deposited magnetar energy $E_{\rm mag}$ and ejected mass \Mej~from our model fits (Figure \ref{fig:mod}; see also \citealt{nic2016b}) we find $v_{\rm sh}\gtrsim5000$\,\kms, consistent with the width of the permitted \ion{O}{1} emission. However, we note that in simple spherical symmetry a dense shell results in a flat-topped line profile, in tension with the sharply-peaked line we observe here. Multi-dimensional effects should be explored, but this is beyond the scope of our study. As this is a recombination line, a further requirement is that some oxygen remains ionised in the inner ejecta (in contrast to the broad [\ion{O}{1}], which is thermally excited and forms further out in the fast, neutral ejecta). The hard radiation field of the engine provides a natural explanation for persistent ionisation.

Using the [\ion{O}{1}]\,$\lambda\lambda6300,6364$ flux in our latest 3 spectra ($\approx3 \times 10^{-15}$\,erg\,s$^{-1}$\,cm$^{-2}$), we can estimate the oxygen mass assuming the line is optically thin \citep{uom1986}:
\begin{equation}
M_{\rm O} = 10^8 \,D^2\,F_{\rm [O I]}\,\exp(22800\,{\rm K}/T),
\label{eq:uom}
\end{equation} 
where $D$ is distance to the SN (529\,Mpc), and $T$ its temperature, which we estimate following \citet{jer2014}: the observed ratio [\ion{O}{1}]\,$\lambda5577$/[\ion{O}{1}]\,$\lambda\lambda6300,6364 = 0.3$ gives $T\approx4900$\,K. This corresponds to $\approx9$\,\M. We find the same result using equation 3 from \citeauthor{jer2014}. \citet{maz2001} found good agreement between oxygen masses derived with equation \ref{eq:uom} and from synthetic spectra, and that oxygen constituted 60--70\% of the ejected mass, giving us a total mass \Mej\,$\approx14$\,\M. This should be treated with caution due to uncertain physical conditions and an exponential temperature dependence: taking the plausible range 4000--6000\,K based on nebular models of hypernovae \citep{maz2001}, the allowed range is $M_{\rm O}=4-25$\,\M, giving \Mej\,$=6-40$\,\M. \citet{jer2016b} have calculated detailed spectroscopic models for SN\,2015bn at +300\,d and find that a similar ejecta mass is required. Interestingly, \citet{maz2016} derived similar masses for other SLSNe using spectroscopic models of the \emph{photospheric} phase; thus a consistent picture is emerging. With our best mass estimate and the observed line velocities ($\approx7500$\,\kms), we infer a kinetic energy $\sim5\times10^{51}$\,erg---higher than that in canonical core-collapse SNe, and closer to hypernovae.

We fit the light curve of SN\,2015bn with a magnetar-powered model in Figure \ref{fig:mod}. We show fits up to +250\,d after peak, and to the full light curve at +400\,d; only small changes in parameters are required (see caption). The parameters are very similar to those from \citet{nic2016b}. With a magnetic field in the range $10^{13}-10^{14}$\,G, we do not expect the SN to be accompanied by a GRB \citep{met2015}, consistent with the radio non-detection from \citet{nic2016b}. The ejecta mass range, \Mej\,$=7-12$\,\M, was calculated assuming an optical opacity $\kappa_{\rm opt}=0.2$\,cm$^2$\,g$^{-1}$. This is appropriate for singly-ionised matter, which is a good approximation around maximum light (oxygen-dominated ejecta at $T\gtrsim10000$\,K). However, after the ejecta cool, the opacity may fall more in line with normal SNe Ic, usually taken to be 0.05--0.1\,cm$^2$\,g$^{-1}$ \citep[e.g.~see][]{ins2013}. This would increase the derived diffusion mass by a factor of 1.6--2.5. Thus we estimate a total plausible range of \Mej\,$=7-30$\,\M, similar to estimates from spectroscopy.

The basic form of the magnetar model eventually follows a $t^{-2}$ decline, which is shallower than the apparently exponential decline we observe. This discrepancy may be a sign of time-variable opacity to the X-rays/$\gamma$-rays from the magnetar (this is distinct from the optical opacity governing the subsequent diffusion of thermalised photons). We demonstrate this with representative fits over the range of $\gamma$-ray opacities studied by \citet{kot2013}, from $\kappa_\gamma=0.01$--0.1\,cm$^2$\,g$^{-1}$, with lower $\kappa_\gamma$ giving a steeper decline. Variable $\kappa_\gamma$ could result from changes in ionisation state, and may be related to the other bumps and wiggles in the light curve \citep{nic2016b}. 

We compare the putative leakage of the magnetar radiation (i.e.~the difference between model and data as a function of time) to the fraction of the total luminosity in emission lines (represented by the strongest [\ion{O}{1}] line). The absolute line flux decreases with time, but the emission relative to the pseudo-continuum gives a measure of just how nebular are the ejecta (see also Figure \ref{fig:cont}). A consistent picture emerges: as the ejecta become more nebular (the optical depth decreases), more of the magnetar radiation escapes without being thermalised, giving a larger deficit in the light curve compared to the model (which assumes an optically thick photosphere).

The deficit at 350--400\,d is $\approx1.4\times10^{42}$\,erg\,s$^{-1}$. If this energy escapes in the X-rays \citep{met2014}, the flux reaching us would be $F_{\rm X}\approx4\times10^{-14}$\,erg\,s$^{-1}$\,cm$^{-2}$ assuming no intervening absorption. Detecting this would provide independent confirmation of the magnetar model. Preliminary analysis of $\sim$15\,ks of Swift-XRT data collected between 2016 July 3--19 puts a limit of $F_{\rm X}<3.6\times 10^{-14}\,\rm{erg\,s^{-1}\,cm^{-2}}$ on the X-ray flux (3$\,\sigma$, 0.3-10 keV, unabsorbed power-law model with index $\Gamma=2$), and thus rules out intrinsic absorption columns $N_{\rm H}<10^{21}\,\rm{cm^{-2}}$ if the excess of energy is effectively leaking out in the 0.3-10 keV energy band. Other observations are planned and will provide tighter constraints to the model. We note that X-rays have been detected in one previous SLSN by \citet{lev2013}. It is also possible that some of the missing flux could be accounted for with mid/far-IR observations.

\section{Conclusions}

We have presented the first comprehensive data (multi-wavelength photometry and multi-epoch spectroscopy) showing a SLSN evolving into the nebular phase. The slow decline of SN\,2015bn would require a radioactive \Ni~mass much larger than the iron mass (and indeed total mass) suggested by the nebular spectrum. The spectrum does not show signs of large \Ni~mass or strong CSM interaction. 
Instead, spectroscopy and light curve fitting suggests an ejected mass $7-30$\,\M, energised by a persistent source of ionising photons, and an inner dense region that could be a signature of overpressure from a central engine.

At $\sim$400\,d, the spectrum is virtually identical to energetic Type Ic SNe such as SNe\,1997dq and 2012au, and extremely similar to GRB-SN\,1998bw. This suggests that the composition and inner density structure are similar, despite their very different luminosities, and confirms the long-held suspicion that SLSNe and GRB-SNe are  related \citep{pas2010,lun2014,nic2015b,gre2015,maz2016}. The similarity to GRB-SNe favours a scenario in which SLSNe result from the engine-powered explosions of stripped massive stars. Scheduled deep X-ray observations at these late epochs may reveal the engine directly.

\acknowledgments

M.N.~thanks the organisers and participants of STScI workshop `The Mysterious Connection Between SLSNe and GRBs' for stimulating discussions, and Carlos Contreras for providing magnitudes to check GMOS flux calibration.
Based on data from ESO as part of PESSTO (188.D-3003, 191.D-0935).
PS1 is supported by NASA Grants NNX12AR65G and NNX14AM74G from NEO Observation Program.
S.J.S.~acknowledges ERC grant 291222
and STFC grants ST/I001123/1 and ST/L000709/1. A.G.Y.~acknowledges ERC grant 307260.
K.M.~acknowledges an Ernest Rutherford Fellowship through STFC.
T.W.C.~acknowledges the Sofia Kovalevskaja Award to P. Schady from the Alexander von Humboldt Foundation of Germany.
S.T.~acknowledges support from TRR33 `The Dark Universe' of the German Research Foundation.

\bibliographystyle{apj}


\begin{table*}
\caption{New data of SN\,2015bn} \label{tab:data}
\begin{tabular}{cccccccc}
\hline
MJD	& Phase  & $u$	&	$g$	&	$r$	&	$i$	&	$z$	&	Telescope \\
\hline
57395.61	&	263.7	&	--		&	--		&	20.60 (0.05)	&	--		&	--		&	PS1\\	
57401.58	&	269.0	&	--		&	21.02 (0.15)	&	20.61 (0.16)	&	20.52 (0.15)	&	20.45 (0.16)	&	LCOGT-2m\\
57403.50	&	270.7	&	--		&	21.09 (0.15)	&	20.77 (0.16)	&	20.63 (0.16)	&	20.46 (0.20)	&	LCOGT-2m\\
57409.47	&	276.1	&	--		&	--		&	20.80 (0.21)	&	20.61 (0.19)	&	20.57 (0.23)	&	LCOGT-2m\\
57423.22	&	288.5	&	22.06 (0.40)	&	21.31 (0.23)	&	21.47 (0.22)	&	21.12 (0.28)	&	20.62 (0.26)	&	NTT+EFOSC2\\
57423.52	&	288.7	&	--		&	--		&	21.31 (0.06)	&	--		&	--		&	PS1\\	
57425.63	&	290.6	&	--		&	21.32 (0.18)	&	21.29 (0.19)	&	21.27 (0.19)	&	--		&	LCOGT-2m\\
57433.55	&	297.7	&	--		&	21.62 (0.20)	&	21.51 (0.21)	&	21.39 (0.19)	&	20.54 (0.22)	&	LCOGT-2m\\
57437.72	&	301.5	&	--		&	21.59 (0.19)	&	21.44 (0.20)	&	21.31 (0.20)	&	20.97 (0.26)	&	LCOGT-2m\\
57444.24	&	307.3	&	22.58 (0.40)	&	21.68 (0.29)	&	21.41 (0.21)	&	-- 	&	21.04 (0.28)	&	NTT+EFOSC2\\
57450.52	&	313.0	&	--		&	21.63 (0.18)	&	21.54 (0.21)	&	21.36 (0.18)	&	--		&	LCOGT-2m\\
57456.27	&	318.1	&	--		&	22.02 (0.23)	&	21.51 (0.25)	&	21.59 (0.20)	&	--		&	NTT+EFOSC2\\
57458.57	&	320.2	&	--		&	21.96 (0.21)	&	21.56 (0.22)	&	21.55 (0.26)	&	--		&	LCOGT-2m\\
57465.53	&	326.4	&	--		&	22.12 (0.24)	&	22.07 (0.23)	&	21.49 (0.23)	&	21.56 (0.26)	&	LCOGT-2m\\
57483.38	&	342.5	&	--		&	--		&	22.09 (0.10)	&	--		&	--		&	PS1\\	
57488.15	&	346.8	&	22.98 (0.30)	&	21.96 (0.22)	&	22.12 (0.21)	&	21.91 (0.17)	&	21.79 (0.29)	&	NTT+EFOSC2\\
\hline
	&  	&	 & $J$	&	$H$	&	$K$	&		&	 \\
\hline	
57404.30	&	271.5	&		&	19.88 (0.28)	&	19.47 (0.32)	&	18.99 (0.34)	&		&	NTT+SOFI\\
57433.21	&	297.4	&		&	20.30 (0.30)	&	20.03 (0.27)	&	19.53 (0.33)	&		&	NTT+SOFI\\
57492.14	&	350.3	&		&	20.72 (0.29)	&	20.83 (0.34)	&	20.20 (0.36)	&		&	NTT+SOFI\\\hline
\hline
MJD    & Phase  & Date & Instrument  & 	Grism or Grating	&  Exposure time (s)	& Airmass    & Average resolution (\AA)\\
\hline			    		                    
57396.3 &  263   & 2016-01-08 &  IMACS       &  G300-17.5  	& 900	&	1.21  &   6  \\
57431.5 &  295   & 2016-02-12 & IMACS       &  G300-17.5  	& 900	&	1.16  &   6  \\
57484.2 &  343   & 2016-04-06 & IMACS       &  G300-17.5  	& 2$\times$900	&	1.25  &   6  \\
57537.8 &  392   & 2016-05-29 & GMOS       &   R150	&  4$\times$1500  	&  	1.11	&	 2  \\

\hline
\end{tabular}


\end{table*}

\end{document}